# A van der Waals pn heterojunction with organic/inorganic semiconductors

Daowei He[1,a)], Yiming Pan[2,a)], Haiyan Nan[3], Shuai Gu[4], Ziyi Yang[1], Bing Wu[1], Xiaoguang Luo[3], Bingchen Xu[1], Yuhan Zhang[1], Yun Li[1], Zhenhua Ni[3], Baigeng Wang[2], Jia Zhu[4], Yang Chai[5], Yi Shi[1,b)] and Xinran Wang[1,b)]

[1]*National Laboratory of Solid State Microstructures, School of Electronic Science and Engineering, and Collaborative Innovation Center of Advanced Microstructures, Nanjing University, Nanjing 210093, China*

[2]*National Laboratory of Solid State Microstructures, School of Physics, Nanjing University, Nanjing 210093, China*

[3]*Department of Physics, Southeast University, Nanjing 211189, China*

[4]*College of Engineering and Applied Science, Nanjing University, Nanjing 210093, China*

[5]*Department of Applied Physics, The Hong Kong Polytechnic University, Hung Hom, Kowloon, Hong Kong, P. R. China*

**Abstract**

**van der Waals (vdW) heterojunctions formed by two-dimensional (2D) materials have attracted tremendous attention due to their excellent electrical/optical properties and device applications. However, current 2D heterojunctions are largely limited to atomic crystals, and hybrid organic/inorganic structures are rarely explored. Here, we fabricate hybrid 2D heterostructures with p-type dioctylbenzothienobenzothiophene ($C_8$-BTBT) and n-type $MoS_2$. We find that**

[a]Daowei He and Yiming. Pan contributed equally to this work.
[b]Corresponding authors. Electronic mail: xrwang@nju.edu.cn; yshi@nju.edu.cn.

**few-layer C$_8$-BTBT molecular crystals can be grown on monolayer MoS$_2$ by vdW epitaxy, with pristine interface and controllable thickness down to monolayer. The operation of the C$_8$-BTBT/MoS$_2$ vertical heterojunction devices is highly tunable by bias and gate voltages between three different regimes: interfacial recombination, tunneling and blocking. The pn junction shows diode-like behavior with rectifying ratio up to 10$^5$ at the room temperature. Our devices also exhibit photovoltaic responses with power conversion efficiency of 0.31% and photoresponsivity of 22mA/W. With wide material combinations, such hybrid 2D structures will offer possibilities for opto-electronic devices that are not possible from individual constituents.**

Heterojunctions are the essential building blocks of modern semiconductor devices such as light-emitting diodes and solid-state lasers.[1,2] With the discovery of graphene and other 2D materials,[3-5] vdW heterojunctions have recently created many attractive opportunities. Unlike the traditional heterojunctions grown by molecular beam epitaxy (MBE), vdW heterojunctions do not require lattice match at the interface, allowing virtually unlimited materials combination. Prototype devices such as tunneling transistors,[6] photodetectors,[7] light-emitting diodes[8,9] and photovoltaic devices[10] have been demonstrated. So far, most of the vdW heterostructures are fabricated by mechanical transfer of 2D layered atomic crystals.[5] Although this method could demonstrate proof-of-concept devices, it cannot be scaled up for real applications. In addition, it is difficult to precisely control the stacking orientation of the heterojunction,[11] which may cause significant variations of device performance.[12]

As an alternative to layered atomic crystals, 2D molecular crystals including oligomers[13] and polymers[14,15] have recently emerged as an interesting class of materials. Particularly, the recent demonstration of vdW epitaxial growth of 2D molecular crystal on graphene suggests the possibility of organic/inorganic hybrid structures.[13,16,17] The epitaxial process offers advantages of low temperature, atomically smooth and clean interface, accurate control of morphology, and the ability to scale up.[13,16-21] The hybrid heterojunctions will further benefit from a much larger library of organic semiconductor materials[22] that allow more design freedom of the devices. So far, however, only limited attempt has been made to interface transition-metal dichalcogenides (TMDs), the most important class of 2D atomic semiconductors, with molecular semiconductors.[23]

In this work, we demonstrated the epitaxial growth of few-layer p-type $C_8$-BTBT molecular crystals on n-type $MoS_2$, and systematically studied the electrical transport and photovoltaic responses of the heterojunction. At room temperature, the pn junction device showed a rectifying ratio up to $10^5$. Under forward bias, the device was operated by either interfacial recombination or tunneling, depending on the backgate voltage. The operation mechanism of the heterojunction device was consistent with band structure analysis and variable-temperature electrical measurements. We also observe strong photovoltaic effects in the heterojunction devices. Our study shows that 2D organic/inorganic vdW heterojunctions may be used for future electronic and optoelectronic device applications.

Recently, we demonstrated the epitaxial growth of highly ordered, few-layer $C_8$-BTBT crystals on graphene and BN. Here we adopted similar methods but with the mechanically exfoliated $MoS_2$ as the epitaxy substrate (see experimental section of supplementary information[24]). Monolayer MoS$_2$ was exfoliated from bulk flakes on 285nm SiO$_2$/Si without thermal annealing and identified by atomic force microscope (AFM) and Raman spectroscopy (Figure 1b, d). Figure 1b and c show the AFM images of the same MoS$_2$ before and after growth, with up to three layers of C$_8$-BTBT grown atop. The C$_8$-BTBT crystals were also confirmed by Raman spectroscopy with two characteristic peaks near 1470cm$^{-1}$ and 1550cm$^{-1}$ (Figure 1d).[13] The growth of C$_8$-BTBT molecular crystals proceeded in a layer-by-layer fashion similar to that on graphene substrate. However, we also observed several interesting distinctions. (1) The thickness of the first C$_8$-BTBT layer (1L) and subsequent layers were ~1.4 nm and ~ 2.9 nm respectively (Figure 1e, S1). This is very different from the growth on graphene, where an additional interfacial layer with thickness ~0.6 nm existed because of the strong molecule-substrate vdW interactions.[13] Since MoS$_2$ is not a π-conjugated system, and the lattice constants are quite different from graphene, the vdW forces between C$_8$-BTBT molecules and MoS$_2$ are significantly reduced and comparable to the inter-molecular interactions. The competition between these forces thus led to the titled molecular packing shown in Figure 1a.[13,16,17] The subsequent layers above 1L were dominated by inter-molecular interactions, giving bulk-like molecular packing[25] with a layer thickness of ~2.9 nm. (2) We observed higher density of nucleation sites forming on MoS$_2$ than on graphene, especially at the edges (Figure S2). Since the nucleation of C$_8$-BTBT occurs preferably at places with high surface energy,[13] we speculate that the high nucleation density may result from surface defects of MoS$_2$. Indeed, many studies have shown that high density of sulfur vacancies, among other defects, exist in $MoS_2$.[26,27] However, the high nucleation density does not appear to significantly affect the vertical charge transport as shown below.

With the $C_8$-BTBT/$MoS_2$ heterostructures, we fabricated vertical field-effect transistors (Figure 2a) and study the electrical transport properties (see Figure S3 for detailed device fabrication procedure[24]). Figure 2c shows the room-temperature current density ($J_{ds} = I_{ds}/A$, where $I_{ds}$ is the source-drain current and $A$ is the area of the heterojunction) as a function of backgate voltage $V_g$ for a representative device ($V_{ds}$=1, 5, and 9V respectively). Interestingly, the transfer characteristics were very different from conventional planar FETs with either $MoS_2$ or $C_8$-BTBT channel (Figure S4). At negative $V_g$, $J_{ds}$ showed a clear peak at low biases. With the increase of $V_g$, $J_{ds}$ abruptly jumped up and becomes roughly independent of $V_g$. However, the conductance in this regime was strongly modulated by $V_{ds}$.

The two regimes in the transfer characteristics suggest different transport mechanisms. In order to understand the operation mechanism, we draw the band diagram of the heterojunction device under forward bias as shown in Figure 2b. Considering that $C_8$-BTBT is a wide bandgap (~3.8 eV) semiconductor with the highest occupied molecular orbit (HOMO) of 5.39eV and lowest unoccupied molecular orbit (LUMO) of 1.55eV,[28] and that the edge of conduction band ($E_{CB}$) and valance band ($E_{VB}$) of monolayer $MoS_2$ are 4.3 eV and 5.9 eV,[29] the heterojunction is of type II with a staggered gap. Since $C_8$-BTBT and $MoS_2$ are p-type and n-type semiconductors as determined by the semiconductor/metal contact, there is only a narrow range in $V_g$ that both materials are conducting (Figure S4). Within this range, electrons and holes are able to inject from the Schottky barriers (SB) at Au/MoS$_2$ and Au/C$_8$-BTBT respectively by thermionic emission (TE), and recombine at the interface (Figure 2b, left panel). Since the conductance in this regime is limited by the minority carrier, a peak in the transfer characteristics is expected (Figure S4). Therefore, we attribute the peak under negative $V_g$ to interfacial recombination regime.

As $V_g$ was further increased beyond the recombination regime, the injection of holes from C$_8$-BTBT was completely blocked, while electrons could still be injected from MoS$_2$. This resulted in strong accumulation of electrons in MoS$_2$ (Figure 2b middle panel). Since the large energy barrier at the C$_8$-BTBT/MoS$_2$ interface blocked thermal activation, the electron transport could only occur via tunneling in this regime. In order to understand the small modulation of tunneling current by $V_g$, we modeled the device as a single-barrier tunnel junction, that is, the barrier at C$_8$-BTBT/MoS$_2$ interface. Under constant $V_{ds}$, the tunneling current is approximately proportional to [6]

$$I_T \propto T(E_F) D_{MoS2}(E_F) \tag{1}$$

where $T(E_F) = \exp\left(-\frac{2d}{\hbar}\sqrt{2m*(U-E_F)}\right)$ is the transmission coefficient through the C$_8$-BTBT layer, $E_F$ is the Fermi energy, $U$=2.75eV is the height of the tunnel barrier, $m*$ is the effective mass, $d$=16nm is the thickness of C$_8$-BTBT layers, $D_{MoS2}(E_F)$ is the density of states (DOS) of MoS$_2$ at the Fermi energy. $D_{MoS2}$ can be modeled as[18]

$$D_{MoS2}(E) = \begin{cases} D_0, & E - E_{CB} \geq 0 \\ D_T \exp(\frac{E-E_{CB}}{\Delta}), & E - E_{CB} < 0 \end{cases} \tag{2}$$

where $D_0$=3.8 × 10$^{14}$eV$^{-1}$cm$^{-2}$ is the DOS in the conduction band of MoS$_2$. The DOS

in the conduction band is a constant because of the 2D nature of $MoS_2$. Below the band edge, the DOS has an exponential tail due to disorders and traps.[18,27] As $V_g$ is swept from negative to positive, $E_F$ of $MoS_2$ is increased proportionally but with very small magnitude because of the large DOS in $MoS_2$. In fact, for a gate overdrive of 40V, $E_F$ only increases by 7.6meV, almost negligible compared $U$. Such small change of $E_F$ (and therefore, transmission coefficient) qualitatively explains the small modulation of tunneling current by $V_g$ as observed experimentally (Figure 2c). Indeed, the calculated $T(E_F)D_{MoS2}(E_F)$ using Equation 1 and 2 clearly captures this feature (Figure 2d). The drop of current below the conduction band edge is due to the decay of DOS in the tail states.

Figure 3a plots the output characteristics under $V_g$=30V, which shows excellent rectifying behavior as expected for a pn junction. The room-temperature rectifying ratio could reach ~$1\times10^5$ (Figure S5b). Under reverse bias, the device showed a blocking behavior with little current flowing because of the increasing SB at both contacts to prevent carrier injection (Figure 3a, inset). Under forward bias, however, we found that the current increased exponentially with $V_{ds}$ under small bias ($V_{ds} < 4$V, Figure S5a) but less dramatically under large bias ($V_{ds} > 6$V). The much weaker current dependence under large bias was due to tunneling dominated process. This was also clear from the band diagram where the SB at the Au/$MoS_2$ contact became thin enough for electrons to tunnel through (Figure 2b, right panel). Under small forward bias, the electron transport at the Au/$MoS_2$ Schottky junction was mainly through TE (Figure 2b, middle panel), leading to the exponential dependence on $V_{ds}$.

In this regime, the output current of the heterojunction device can be calculated as

$$I_{tot} \propto \text{T}(V_g) \times \text{I}_{\text{TE}} = I_0 \left( exp\left(\frac{eV_{ds}}{\eta k_B T}\right) - 1 \right) \quad (3)$$

where $T(V_g)$ is the tunneling transmission coefficient through $C_8$-BTBT, $\eta$ is the ideality factor considering the fact that $V_{ds}$ does not fully drop on the Schottky junction,

$$I_0 = \text{T}(V_g) A A^* T^2 e^{-e\Phi_{SB}/k_B T} \quad (4)$$

is the reverse saturation current, $A$ is the area of the Schottky junction, $A^*$ is the effective Richardson constant, and $\Phi_{SB}$ is the SB height for electrons at the Au/MoS$_2$ contact. From variable-temperature measurements, we were able to extract $\Phi_{SB}$, which is an important device parameter. Figure 3b shows the output characteristics under 180K, 240K and 300K respectively. We could fit all the output characteristics with the same ideality factor η = 20.6 (Figure S6) and extract the reverse saturation current $J_0 = I_0/A$ as a function of temperature (Figure 3b inset, symbols). The $J_0$-$T$ relationship is well described by Equation 4 with $\Phi_{SB}$=120meV (Figure 3b inset, line). The extracted $\Phi_{SB}$ is consistent with the widely observed n-type behavior in Au-contacted MoS$_2$ transistors.[30] The small SB for electrons suggests strong Fermi level pinning at the Au/MoS$_2$ interface,[31,32] likely dominated by defects and interfacial traps in MoS$_2$.

We were also able to corroborate the proposed device model by low temperature measurements. Figure 3c is the Arrhenius plot of current density under two different regimes. For $V_{ds}$ > 6V, we observed that the conductance was insensitive to temperature, a strong evidence for tunneling-dominated current. However, for $V_{ds}$ <

4V, the conductance showed an exponential relationship with temperature, consistent with TE. From the linear fitting of the Arrhenius plot (Figure 3c, blue dashed line), $\Phi_{SB}$ was extracted and found to scale linearly with $V_{ds}^{1/2}$, due to the image force as in conventional Schottky junctions.[2] The extrapolated $\Phi_{SB}$=190meV at zero bias (Figure 3c inset) is in good agreement with the fitting in Figure 3, reassuring the consistency of our theoretical model.

We further investigated the photo-response of our devices. To this end, we carried out photovoltaic measurements under the white light illumination from a standard solar simulator with incident optical power $P_{opt}$ varied between 100 and 1100 W/m$^2$ in ambient condition. Figure 4 shows clear photovoltaic effect of a representative device with 8-layer C$_8$-BTBT crystals. The open circuit voltage $V_{oc}$ is about 0.5V. The power conversion efficiency, defined as $\eta = P_{el,m}/P_{opt}$, is up to 0.31%, which is comparable to values reported for lateral monolayer WSe$_2$ pn junctions.[33] The photo responsibility $R = I_{ds}/P_{opt}$ under zero $V_{ds}$ is ~22mA/W. We make it clear that the power conversion efficiency is estimated using the exposed area of the heterojunction, rather than the area covered by Au. We also note that the efficiency is just a rough estimate. We believe much better performances are possible with further device optimization (e. g. transparent top electrodes). The photo-response should be a general feature of the hybrid heterojunction. With many possible combinations of TMDs and organic materials, we expect this type of devices could be very versatile and useful in photovoltaic or photodetector applications.

In conclusion, we have demonstrated that high-quality few-layer molecular semiconductors can be epitaxially grown on TMDs, creating a 2D hybrid organic/inorganic vdW heterojunction. In a vertical heterojunction device created by p-type $C_8$-BTBT and n-type $MoS_2$, we observed excellent rectifying behavior, as well as strong photovoltaic responses. Considering the huge library of organic semiconductors, our work opens up many design possibilities for 2D heterostructure devices.

**Acknowledgements.** We thank NIPPON KAYAKU Co., Ltd. Japan for providing $C_8$-BTBT materials. This work was supported in part by National Key Basic Research Program of China 2013CBA01604, 2015CB921600; National Natural Science Foundation of China 61325020, 61261160499, 11274154, 61521001, 61574074; MICM Laboratory Foundation 9140C140105140C14070; a project funded by the Priority Academic Program Development of Jiangsu Higher Education Institutions; "Jiangsu Shuangchuang" program and 'Jiangsu Shuangchuang Team' Program.

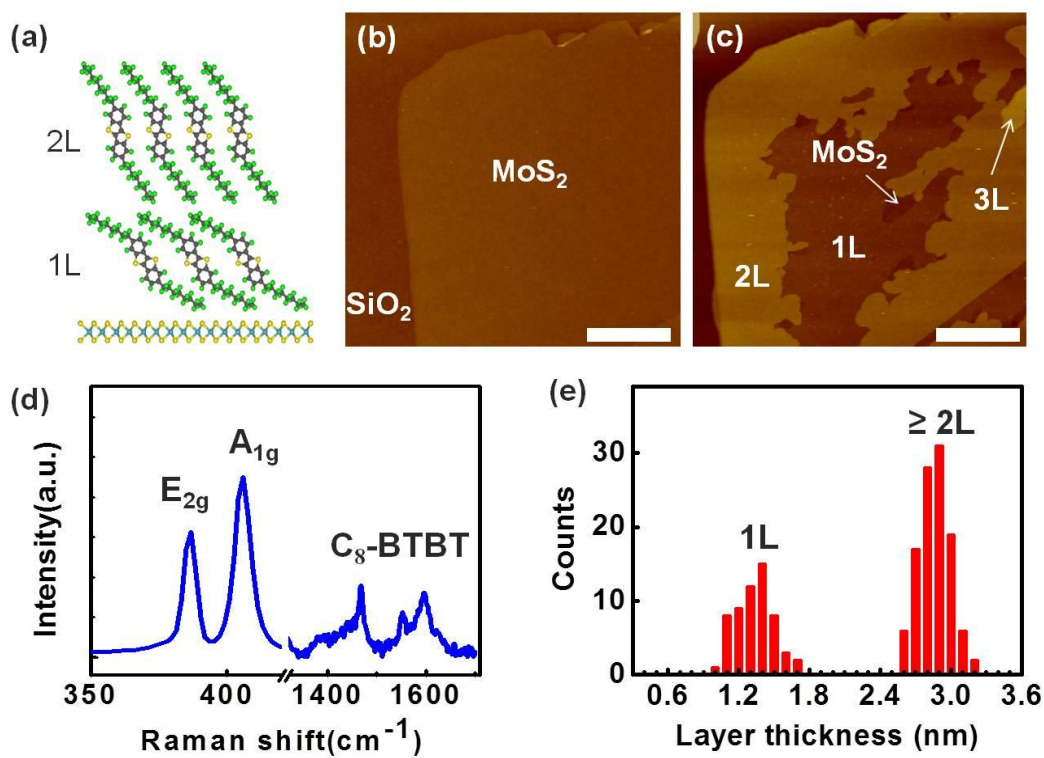

**Figure 1.** (a) A schematic illustration of molecular packing of $C_8$-BTBT on $MoS_2$. (b,c) AFM snapshots of $MoS_2$ before and after growth of $C_8$-BTBT molecular crystal. Scale bars, 1μm. (d) Raman spectrum of $C_8$-BTBT grown on $MoS_2$. The $E_{2g}$ and $A_{1g}$ peaks are from $MoS_2$, and the other two peaks near 1470 cm$^{-1}$ and 1550 cm$^{-1}$ are from $C_8$-BTBT. (e) Histogram of the layer thickness of $C_8$-BTBT molecular crystals on $MoS_2$, from over 5 samples.

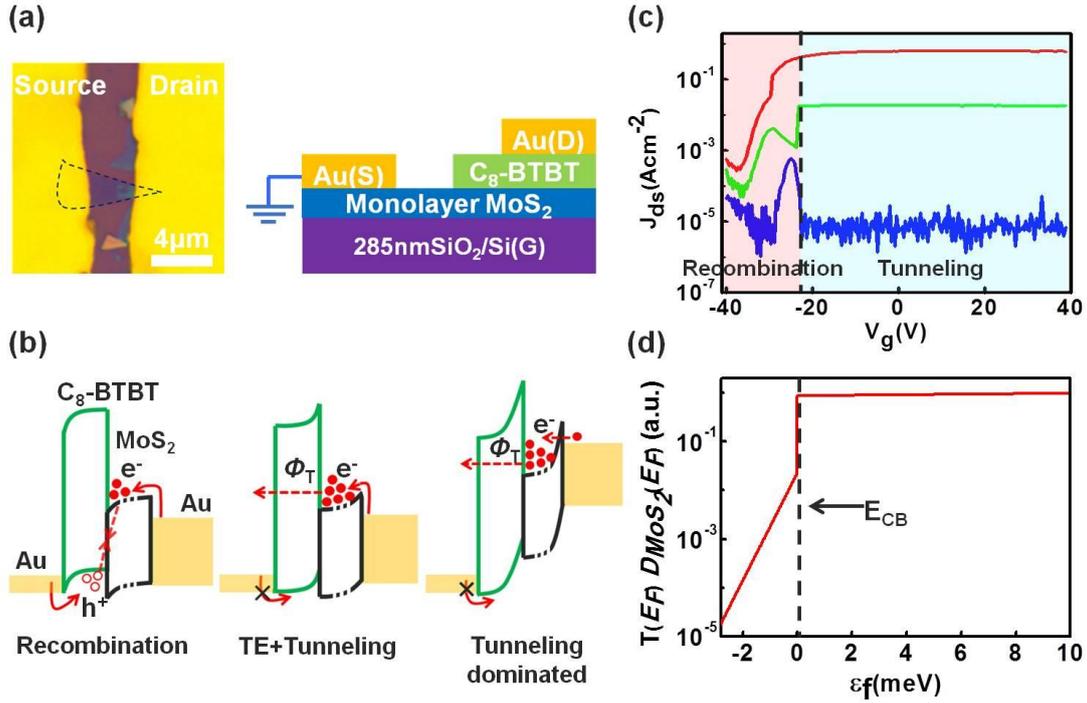

**Figure 2.** (a) Left, microscope image of a vertical heterojunction device. The dotted line represents the shape of the monolayer $MoS_2$. The source and drain electrode is in contact with $MoS_2$ and $C_8$-BTBT respectively. The multi-layer $C_8$-BTBT can be observed near the edge of the drain electrode. Right, schematic illustration of the device. (b) The band diagram of $MoS_2/C_8$-BTBT vertical heterojunction device under forward bias, showing different operation regimes. The dotted line is used to represent the lateral part of $MoS_2$. (c) Room-temperature $J_{ds}$-$V_g$ characteristics of a typical $MoS_2/C_8$-BTBT vertical heterojunction device, $V_{ds}$=1V(blue line), 5V(green line) and 9V(red line). (d) The calculated $T(E_F)D_{MoS2}(E_F)$ as a function of $E_F$ in the tunneling regime. During the calculation, we used the following parameters: $D_T$=$10^{13}$eV$^{-1}$cm$^{-2}$, $\Delta$=14meV. The dashed vertical line represents the conduction band edge of $MoS_2$.

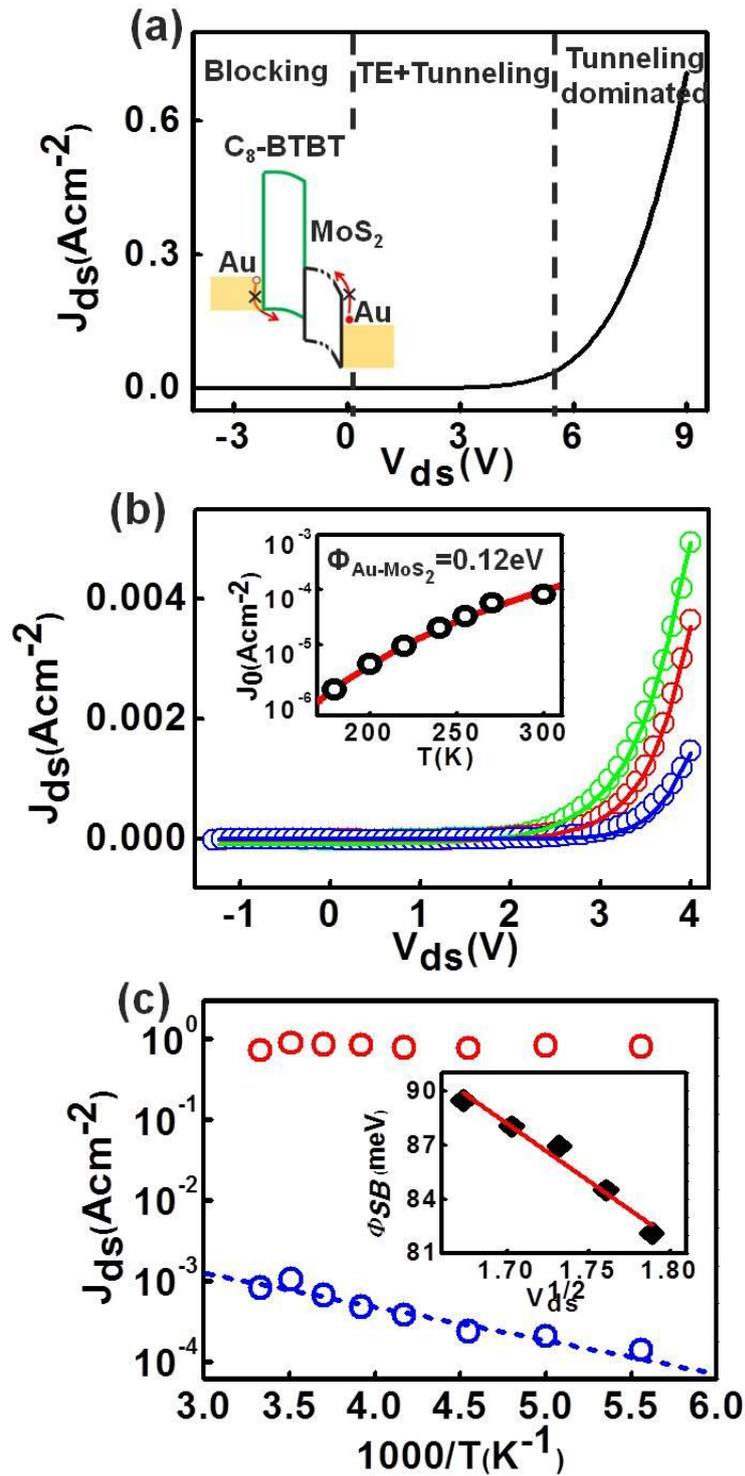

**Figure 3.** (a) Room-temperature $J_{ds}$-$V_{ds}$ characteristics of same device in Figure 2 under $V_g$=30V. The different operation regimes are separated by vertical dashed lines. Inset shows the band diagram under reverse bias. (b) $J_{ds}$-$V_{ds}$ characteristics of same device in (a) under $V_g$=30V at T=300K (red symbols), 240K (green symbols) and 180K (blue symbols). Lines are the fitting results using Equation 2. Inset: the reverse saturation current density $J_0$ as a function of temperature (black symbols) and theoretical fitting using Equation 4 (red line). $\Phi_{SB}=0.12$eV is derived for the Au/MoS$_2$ interface. (c) Arrhenius plot of $J_{ds}$ of the same device in (a) under $V_g=30$V. Red symbol, $V_{ds}=9$V, blue symbol, $V_{ds}=3$V. The blue dashed line is the linear fitting of the Arrhenius plot. The extracted $\Phi_{Au-MoS2}$ are plotted in the inset as a function of $V_{ds}^{1/2}$ (black square). The extrapolated $\Phi_{Au-MoS2}$ at zero bias is 0.19 eV (red line).

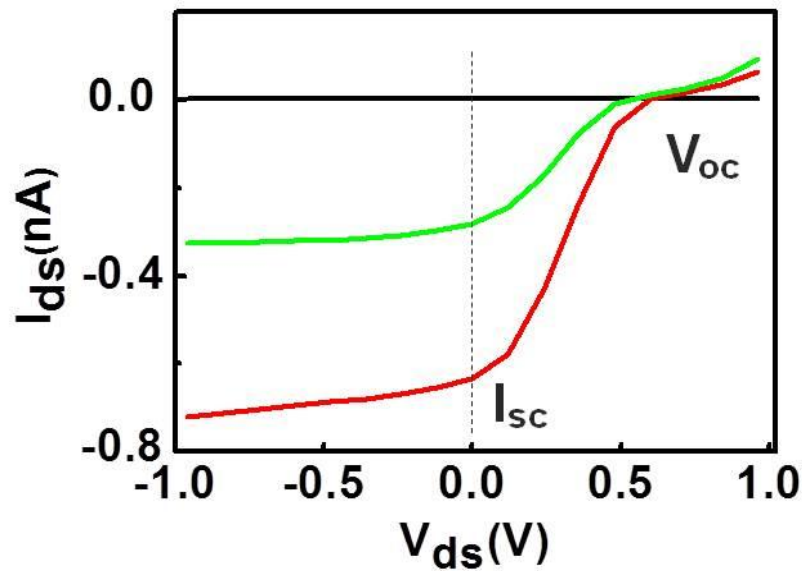

**Figure.4.** $I_{ds}-V_{ds}$ characteristics of a heterojunction device under white light illumination with $P_{opt}=1100$W/m$^2$ (red), 550 W/m$^2$ (green) and under dark condition (black). The measurements were taken at $V_g=-30$V under ambient condition. $V_{OC}$ and $I_{SC}$ present open circuit voltage and short circuit current respectively.

# A van der Waals pn heterojunction with organic/inorganic semiconductors

Daowei He[1,a)], Yiming Pan[2,a)], Haiyan Nan[3], Shuai Gu[4], Ziyi Yang[1], Bing Wu[1], Xiaoguang Luo[3], Bingchen Xu[1], Yuhan Zhang[1], Yun Li[1], Zhenhua Ni[3], Baigeng Wang[2], Jia Zhu[4], Yang Chai[5], Yi Shi[1,b)] and Xinran Wang[1,b)]

[1]*National Laboratory of Solid State Microstructures, School of Electronic Science and Engineering, and Collaborative Innovation Center of Advanced Microstructures, Nanjing University, Nanjing 210093, China*

[2]*National Laboratory of Solid State Microstructures, School of Physics, Nanjing University, Nanjing 210093, China*

[3]*Department of Physics, Southeast University, Nanjing 211189, China*

[4]*College of Engineering and Applied Science, Nanjing University, Nanjing 210093, China*

[5]*Department of Applied Physics, The Hong Kong Polytechnic University, Hung Hom, Kowloon, Hong Kong, P. R. China*

## Supplementary Information

**Experimental Section**

*Growth of $C_8$-BTBT crystals.* The vdW epitaxial growth was carried out in home-built tube furnace similar to Reference 13. We put the $C_8$-BTBT powder (supplied by NIPPON KAYAKU Co., Ltd.) into the center of quartz tube, and the substrate was placed several centimeters away from the source. We then evacuated the quartz tube to ~$4\times10^{-6}$ torr and heated up the $C_8$-BTBT powder to $130^{\circ}$C to start growth. We could select the best crystallization zone of the $C_8$-BTBT molecular crystal by changing the sample location in the tube. The $C_8$-BTBT molecular crystal could be precisely controlled by the source temperature, growth time, and sample location.

*Characterization of the $C_8$-BTBT crystals.* The AFM was performed on a Veeco Multimode 8 under ambient conditions, with soft taping mode. Micro-Raman spectroscopy was performed on a LabRAM HR800 The excitation laser was 532nm with power of 1mW and spot size ~ 1μm .

*Electrical measurements.* Variable-temperature electrical measurements were performed by an Agileng B1500 semiconductor parameter analyzer in a Lakeshore low-temperature probe station with base pressure ~ $10^{-5}$ Torr. No annealing steps were necessary before electrical measurements.

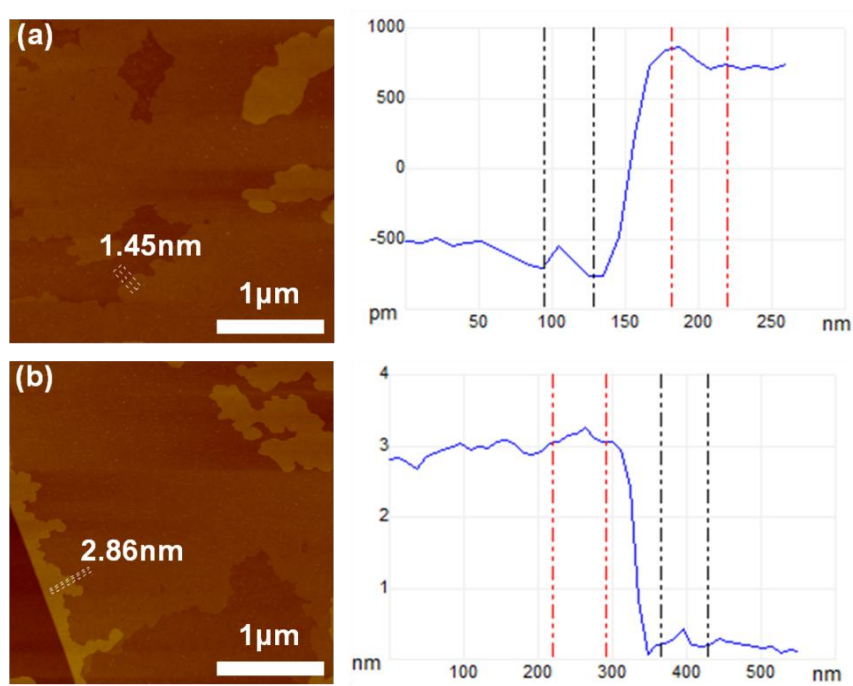

**FIG. S1.** Thickness measurement of 1L (a) and 2L (b) $C_8$-BTBT molecular crystals grown on $MoS_2$. The thickness of the 1L and 2L sample is ~1.45nm and 2.86nm, respectively.

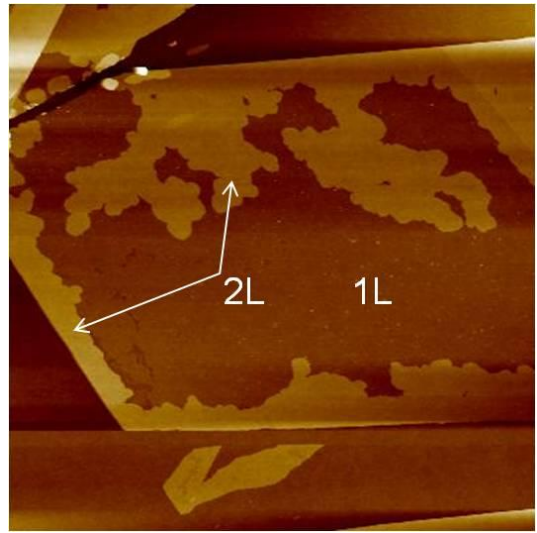

**FIG. S2.** Another sample of two layers $C_8$-BTBT molecular crystals grown on $MoS_2$. The 1L $C_8$-BTBT film has completely covered the $MoS_2$, while 2L has partially grown on 1L $C_8$-BTBT.

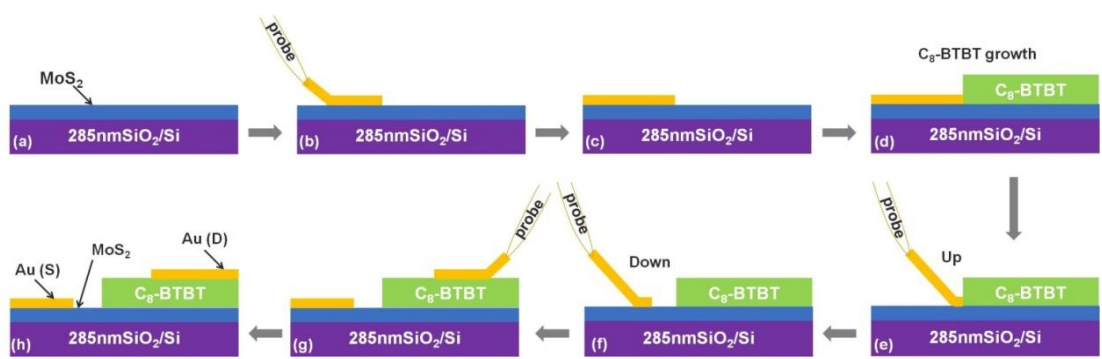

**FIG. S3.** The fabrication process of vertical $MoS_2$/$C_8$-BTBT heterojunction field-effect transistor. We first exfoliated monolayer $MoS_2$ on silicon substrate with 285nm thermal oxide as a growth substrate without thermal annealing (a). Then we transferred a 100nm-thick Au electrode to cover part of the MoS$_2$ during the C$_8$-BTBT growth (b, c), and few-layer C$_8$-BTBT molecular crystals are grown on MoS$_2$ in a tube furnace (d). After growth, we moved the Au electrode backward several microns as the source (ground) electrode (e, f). The moving of electrode was to create a gap between the source electrode and C$_8$-BTBT layer (Figure 2a). Finally, we transferred another Au electrode onto the C$_8$-BTBT molecular film as drain electrode (g, h). All the transfer and manipulation steps of Au electrode were performed under an optical microscope using a tungsten probe tip attached to a micro manipulator.

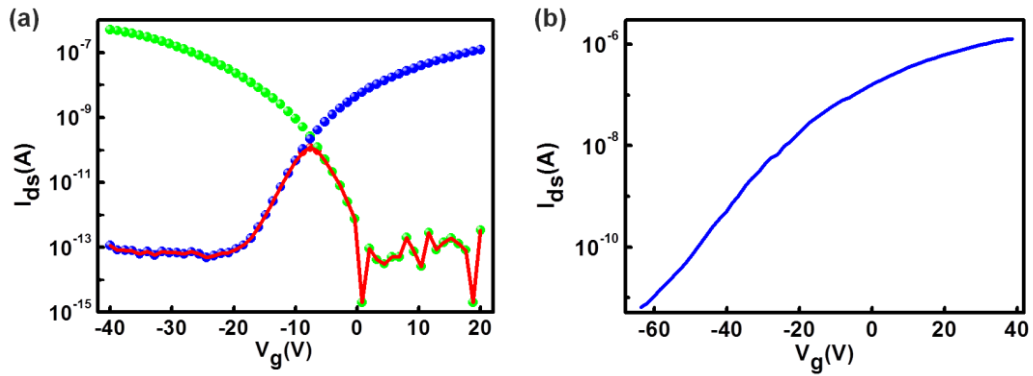

**FIG. S4.** (a) Room-temperature $I_{ds}$-$V_g$ characteristics of planar MoS$_2$ (blue symbol, $V_{ds}$=0.1V) and C$_8$-BTBT (green symbols, $V_{ds}$=0.5V) field-effect transistors. There is only a limited $V_g$ range (-20V ~ 0V) that both devices are turned on. The red line plots the $I_{ds} = \frac{I_{ds(MoS2)} \times I_{ds(C8-BTBT)}}{I_{ds(MoS2)} + I_{ds(C8-BTBT)}}$, showing that a peak is expected for recombination regime. (b) $I_{ds}$-$V_g$ characteristics of another back-gated monolayer MoS$_2$ FETs on SiO$_2$ substrate. $V_{ds}$=100mV. This device was not completely turned off even at $V_g$< *-40V*. So it's possible to have conduction peaks at $V_g$< *-20V* as shown in the main text.

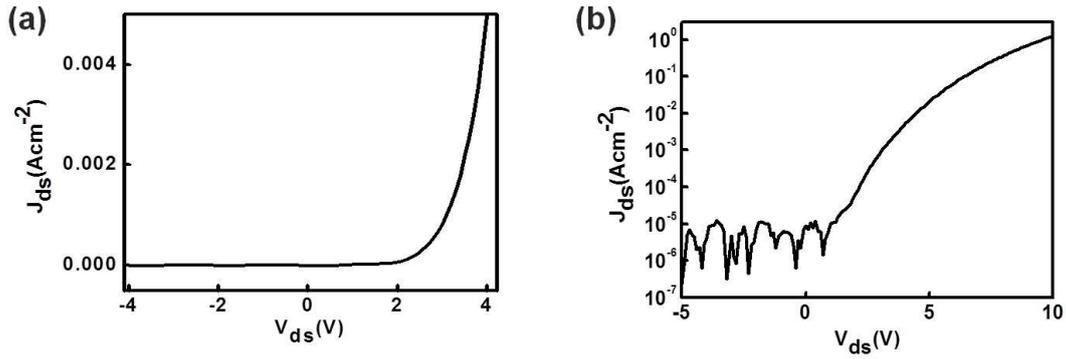

**FIG. S5.** a) The $J_{ds}$-$V_g$ characteristics of same device in Figure 2 under small bias. The current increase exponentially within the voltage at small bias range ($V_{ds}$<4V), consistent with a rectification characteristic of the p-n junction. b) The $J_{ds}$-$V_g$ characteristics of same device in log scale. The device exhibits excellent rectification characteristics, the room-temperature rectifying ratio could reach up to ~$1 \times 10^5$.

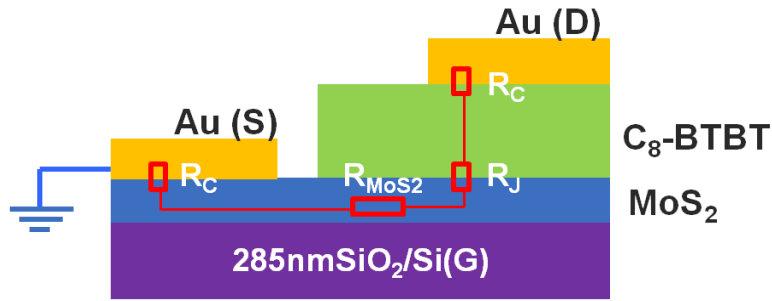

**FIG. S6.** The current conducting path and associated resistance of the device. $R_{MoS2}$ is the resistance of monolayer $MoS_2$ sheet not covered by $C_8$-BTBT, $R_c$ is the contact resistance including the effect of SB, $R_J$ is the resistance of the tunnel junction. Under forward bias, only a fraction of bias voltage drops on the Schottky junction at Au/$MoS_2$ interface, resulting in a large ideality factor as determined from the fitting in Figure 3b.